# Experimental Modeling of Cosmological Inflation with Metamaterials

Igor I. Smolyaninov, Yu-Ju Hung, Ehren Hwang

*Department of Electrical and Computer Engineering, University of Maryland, College Park, MD 20742, USA*

*phone: 301-405-3255;  fax: 301-314-9281;  e-mail: smoly@umd.edu*

**Recently we demonstrated that mapping of monochromatic extraordinary light distribution in a hyperbolic metamaterial along some spatial direction may model the flow of time and create an experimental toy model of the big bang. Here we extend this model to emulate cosmological inflation. This idea is illustrated in experiments performed with two-dimensional plasmonic hyperbolic metamaterials. Spatial dispersion which is always present in hyperbolic metamaterials results in scale-dependent (fractal) structure of the inflationary "metamaterial spacetime". This feature of our model replicates hypothesized fractal structure of the real observable universe.**

### I. Introduction

Electromagnetic metamaterials are capable of emulating many interesting space-time geometries [1]. For example, very recently it was demonstrated that extraordinary rays in a hyperbolic metamaterial behave as particle world lines in a three dimensional (2+1) Minkowski spacetime [2,3]. When this spacetime is "curved", metamaterial analogues of black holes [4] and the big bang [3] may be created. Moreover, recent developments in quantum chromodynamics [5] and electromagnetism of metamaterials [6] strongly indicate a direct connection between the microscopic physics of space-time and the physics of hyperbolic metamaterials. It appears that physical vacuum subjected to very strong magnetic field behaves as a hyperbolic metamaterial. Therefore, electromagnetic models of various exotic spacetimes appear to be quite useful in understanding microscopic behavior of physical vacuum. Maxwell equations, which describe light propagation in metamaterials may be made exactly the same as equations which describe particle propagation in some particular spacetime geometry. Therefore, essential physics must be similar. On the other hand, this analogy is as good as the equations themselves. While we are sure that Maxwell equations describing metamaterials are correct, we do not know if the general relativity description of real astrophysical black holes is valid. Detailed study of metamaterial models may highlight potential differences. For example, it is believed that the properties of physical spacetime on the Planck scale may be quite unusual. Spacetime may become quantized resulting in the speed of light being dependent on photon energy [7]. In the language of optical physics, this effect is called dispersion. Modeling of dispersion with metamaterials is very easy. The goal of this paper is to extend recently introduced metamaterial model of the big bang [3] to emulate cosmological inflation. It appears that such an extension is necessarily affected by strong spatial dispersion, which is always present in hyperbolic metamaterials. Effects of spatial dispersion result in scale-dependent (fractal) structure of the inflationary "metamaterial spacetime" in our model. We note that this feature of our model replicates hypothesized fractal structure of the real observable universe [8,9].

Over the last few decades inflation has become the main paradigm in cosmology [10]. It is characterized by extremely fast exponential expansion of space during a brief period of time immediately after the big bang. As a result of this expansion, all observable universe originated in a small causally connected region, which explains the fact that the observable universe appears flat, homogeneous, and isotropic. On the other hand, this fast expansion necessarily leads to formation of the cosmological horizon: due to exponential expansion of space, two nearby observers are separated very fast, so that the distance between them quickly exceeds the limits of communications. Therefore, these observers eventually end up in different causal patches (universes) of the de Sitter "multiverse". In order to understand how we can create an experimental model of cosmological inflation, let us first recall the main features of the metamaterial model of the big bang, which is described in detail in ref. [3].

Let us assume that a non-magnetic uniaxial anisotropic metamaterial has constant dielectric permittivities $\varepsilon_x = \varepsilon_y = \varepsilon_1 > 0$ and $\varepsilon_z = \varepsilon_2 < 0$ in some frequency range around $\omega = \omega_0$. Any electromagnetic field propagating in this metamaterial can be expressed as a sum of the "ordinary" ($\vec{E}$ perpendicular to the optical axis) and "extraordinary" ($\vec{E}$ parallel to the plane defined by the k–vector of the wave and the optical axis) contributions. We will consider the extraordinary component of the field and introduce a "scalar" wave function as $\varphi = E_z$. Since metamaterials exhibit large dispersion, we will work in the frequency domain and write the macroscopic Maxwell equations as

$$\frac{\omega^2}{c^2}\vec{D}_\omega = \vec{\nabla}\times\vec{\nabla}\times\vec{E}_\omega \text{ and } \vec{D}_\omega = \vec{\vec{\varepsilon}}_\omega \vec{E}_\omega \qquad (1)$$

Let us assume that the metamaterial is illuminated by coherent CW laser field at frequency $\omega_0$, and we study spatial distribution of the extraordinary field $\varphi_\omega$ at this frequency. Eq.(1) results in the following wave equation for $\varphi_\omega$:

$$-\frac{\partial^2 \varphi_\omega}{\varepsilon_1 \partial z^2} + \frac{1}{|\varepsilon_2|}\left(\frac{\partial^2 \varphi_\omega}{\partial x^2} + \frac{\partial^2 \varphi_\omega}{\partial y^2}\right) = \frac{\omega_0^2}{c^2}\varphi_\omega = \frac{m^{*2}c^2}{\hbar^2}\varphi_\omega \qquad (2)$$

This equation looks like the 3D Klein-Gordon equation describing a massive scalar $\varphi_\omega$ field in which the spatial coordinate $z=\tau$ behaves as a "timelike" variable. Therefore, eq.(2) describes world lines of massive particles which propagate in a flat (2+1) Minkowski spacetime. If $\varepsilon_1$ and $\varepsilon_2$ are allowed to vary in coordinate space, eq.(2) will remain approximately valid if $\partial\varepsilon/\partial x^i << k_i$. Therefore, the Klein-Gordon equation for a massive particle in a gravitational field [11]

$$\frac{1}{\sqrt{-g}}\frac{\partial}{\partial x^i}\left(g^{ik}\sqrt{-g}\frac{\partial\varphi}{\partial x^k}\right)=\frac{m^2c^2}{\hbar^2}\varphi \qquad (3)$$

can be emulated. For example, let us consider an experimental situation, in which we allow variation of $\varepsilon_2$ as a function of $z$, while $\varepsilon_1$ is kept constant. According to eqs.(2,3) this situation corresponds to "cosmological expansion" of the (2+1) dimensional universe as a function of "timelike" $z=\tau$ variable. A simple experimental model of the big bang based on this idea has been demonstrated in ref.[3] using plasmonic hyperbolic metamaterial. A summary of this experiment is presented in Fig.1.

**II. Results**

Let us extend this approach to demonstration of cosmological inflation. The metric of (2+1) dimensional inflationary de Sitter spacetime (Fig.2(a)) can be written as [10]

$$ds^2 = -dt^2 + e^{Ht}(dx^2+dy^2) \qquad (4)$$

where the Hubble constant $H\sim\Lambda^{1/2}$ ($\Lambda$ is the cosmological constant). The corresponding Klein-Gordon equation is

$$-\frac{\partial^2\varphi}{\partial t^2}+\frac{1}{e^{Ht}}\left(\frac{\partial^2\varphi}{\partial x^2}+\frac{\partial^2\varphi}{\partial y^2}\right)-H\left(\frac{\partial\varphi}{\partial t}\right)=\frac{m^2c^2}{\hbar^2}\varphi \qquad (5)$$

Now let us consider extraordinary field in a hyperbolic metamaterial having z-dependent $\varepsilon_1=\varepsilon_x=\varepsilon_y>0$ and $\varepsilon_2=\varepsilon_z<0$. Taking into account z derivatives of $\varepsilon_1$ and $\varepsilon_2$, eq.(1) results in the following equation for $\varphi_\omega$:

$$-\frac{\partial^2 \varphi_\omega}{\varepsilon_1 \partial z^2}+\frac{1}{(-\varepsilon_2)}\left(\frac{\partial^2 \varphi_\omega}{\partial x^2}+\frac{\partial^2 \varphi_\omega}{\partial y^2}\right)+\left(\frac{1}{\varepsilon_1^2}\left(\frac{\partial \varepsilon_1}{\partial z}\right)-\frac{2}{\varepsilon_1 \varepsilon_2}\left(\frac{\partial \varepsilon_2}{\partial z}\right)\right)\left(\frac{\partial \varphi_\omega}{\partial z}\right)+$$
$$+\frac{\varphi_\omega}{\varepsilon_1 \varepsilon_2}\left(\frac{1}{\varepsilon_1}\left(\frac{\partial \varepsilon_1}{\partial z}\right)\left(\frac{\partial \varepsilon_2}{\partial z}\right)-\left(\frac{\partial^2 \varepsilon_2}{\partial z^2}\right)\right)=\frac{\omega_0^2}{c^2}\varphi_\omega \qquad (6)$$

It is easy to see that $\varepsilon_2=const<0$ and $\varepsilon_1 \sim e^{-Hz}$ (where $z=\tau$ is considered a time-like variable) differs from eq.(5) only by the scaling factor $\varepsilon_2$ in the xy-direction in the limit of large $Hz$:

$$-\frac{\partial^2 \varphi_\omega}{\partial z^2}+\frac{1}{(-\varepsilon_2)e^{Hz}}\left(\frac{\partial^2 \varphi_\omega}{\partial x^2}+\frac{\partial^2 \varphi_\omega}{\partial y^2}\right)-H\left(\frac{\partial \varphi_\omega}{\partial z}\right)=\frac{\omega_0^2}{c^2 e^{Hz}}\varphi_\omega \approx 0 \qquad (7)$$

In this limit eq. (7) describes propagation of massless particles in the inflationary de Sitter metric described by eq.(4). On the other hand, choosing $\varepsilon_1=const>0$ and $\varepsilon_2 \sim -e^{Hz}$ reproduces eq.(5) for massive particles if we introduce new wave function $\psi$ as $\psi=(-\varepsilon_2)^{1/2}\varphi_\omega$. Indeed, such a substitution leads to the following wave equation for $\psi$:

$$-\frac{\partial^2 \psi}{\partial z^2}+\frac{\varepsilon_1}{e^{Hz}}\left(\frac{\partial^2 \psi}{\partial x^2}+\frac{\partial^2 \psi}{\partial y^2}\right)-H\left(\frac{\partial \psi}{\partial z}\right)=\left(\frac{\varepsilon_1 \omega_0^2}{c^2}+\frac{H^2}{4}\right)\psi \qquad (8)$$

which differs from eq.(5) only by the scaling factor $\varepsilon_1$ in the xy-direction, and is valid at any z. Thus, both metamaterial choices are suitable for modeling cosmological inflation in the lab. The former choice also clearly illustrates the effect of spatial dispersion on the metamaterial models of cosmological inflation.

6Let us recall that in addition to frequency dispersion, hyperbolic metamaterials typically exhibit spatial dispersion, which means that $\varepsilon_1$ and $\varepsilon_2$ exhibit weak dependence on the wave vector of the form

$$\varepsilon = \varepsilon^{(0)} + \beta\left(\frac{k^2 c^2}{\omega^2}\right) \qquad (9)$$

where $\beta$ is small [12]. When the absolute values of $\varepsilon^{(0)}_1$ and $\varepsilon^{(0)}_2$ are large, spatial dispersion can be disregarded, and description of the hyperbolic medium in terms of effective (2+1) Minkowski spacetime is a good approximation. On the other hand, the effects of spatial dispersion may become strong if either $\varepsilon_1$ or $\varepsilon_2$ are close to zero. In the particular case considered above (in which $\varepsilon_2$=const<0 and $\varepsilon_1 \sim e^{-Hz}$, where $z=\tau$ is considered a time-like variable) $\varepsilon_1$ is indeed small so that the effects of spatial dispersion must be taken into account. Examination of eqs.(6,7) demonstrates that the length elements of the effective "metamaterial space" can be written as

$$dl^2 = -\frac{\varepsilon_2}{\varepsilon_1}\left(dx^2 + dy^2\right) = -\frac{\varepsilon_2}{e^{-Hz} + \beta\frac{k^2 c^2}{\omega_0^2}}\left(dx^2 + dy^2\right), \qquad (10)$$

which indicates that the measured length intervals are strongly scale-dependent in the limit of large $Hz$. This feature of the metamaterial model replicates hypothesized fractal structure of the real observable universe [8,9].

Let us consider a possible experimental realization of the inflationary universe using the 3D layered hyperbolic metamaterial structure shown in Fig.2(b). Let us assume that the metallic layers are oriented perpendicular to $z=\tau$ direction. The diagonal components of the permittivity tensor in this case have been calculated in ref. [13] using Maxwell-Garnett approximation:

$$\varepsilon_1 = \alpha\varepsilon_m + (1-\alpha)\varepsilon_d \quad , \quad \varepsilon_2 = \frac{\varepsilon_m\varepsilon_d}{(1-\alpha)\varepsilon_m + \alpha\varepsilon_d} \qquad (11)$$

where $\alpha$ is the fraction of metallic phase, and $\varepsilon_m<0$ and $\varepsilon_d>0$ are the dielectric permittivities of metal and dielectric layers, respectively. We would like to produce the following behavior of anisotropic permittivity during "inflation": $\varepsilon_2 \sim$ const$<0$ and $\varepsilon_1 \sim e^{-Hz}$ by changing $\alpha$ as a function of $z$. Simple analysis of eqs.(11) indicates that

$$\alpha = \frac{\varepsilon_d - e^{-Hz}}{\varepsilon_d - \varepsilon_m} \qquad (12)$$

produces the required functional dependence of $\varepsilon_1 \sim e^{-Hz}$, while

$$\varepsilon_2 = \frac{\varepsilon_m\varepsilon_d}{\varepsilon_m + \varepsilon_d - e^{-Hz}} \approx const < 0 \qquad (13)$$

condition will be satisfied if $\varepsilon_d > -\varepsilon_m$. Therefore, such low loss "alternative plasmonic materials" [14] as indium tin oxide (ITO) can be used as the "metallic" component of the required layered hyperbolic metamaterial in the near IR (1.5-2 μm) spectral range. Thus, by gradually increasing $\alpha$ towards the $\alpha_0=\varepsilon_d/(\varepsilon_d-\varepsilon_m)$ value we will observe the effect in question. This can be achieved by gradual increase of the metal layer thickness $d_1$, while keeping the dielectric layer thickness $d_2$ constant, so that the necessary range of $\alpha=d_1/(d_1+d_2)$ is achieved. Let us note that even more interesting realization of cosmological inflation may be achieved using a layered hyperbolic metamaterial based on nonlinear dielectric layers having negative nonlinear susceptibility $\chi^{(3)}$. In such a case the required behaviour of $\varepsilon_1$ near zero may be achieved by self-focusing of the optical field without any gradient of $\alpha$.

While experimental demonstration of the "inflationary universe" with 3D metamaterials would require sophisticated nanofabrication, experimental demonstration of this concept using PMMA-based plasmonic hyperbolic metamaterials described in detail in ref.[3] is much simpler. The world line behavior near the big bang (Fig.1(c)) was emulated in a 2D hyperbolic metamaterial having circular symmetry with constant $\varepsilon_\theta>0$ and $\varepsilon_r<0$, so that the wave equation is

$$-\frac{\partial^2 \varphi_\omega}{\varepsilon_\theta \partial r^2}+\frac{1}{|\varepsilon_r|}\frac{\partial^2 \varphi_\omega}{r^2 \partial \theta^2}=\frac{m^{*2}c^2}{\hbar^2}\varphi_\omega \quad , \qquad (14)$$

and $r=\tau$ may be considered as a timelike variable. Plasmon rays were launched into the hyperbolic metamaterial near $r=0$ point via the central phase matching structure marked with an arrow in Figs.1(b,c). Similar to the world line behavior near the big bang, which is shown in Fig.1(a), plasmonic rays or "world lines" indeed increase their spatial separation as a function of "timelike" radial coordinate $r=\tau$. The point (or moment) $r=\tau=0$ corresponds to a moment of the toy "big bang". Let us describe how this experimental model may be extended in order to emulate the inflationary stage of universe expansion shortly after the big bang.

Plasmonic field propagation far from the center of the concentric metamaterial structure may be described locally using rectangular coordinates so that radial coordinate $r$ corresponds to $z$, and angular coordinate $\theta$ corresponds to $x$ in the metamaterial model described by eq.(7). Thus, we need to produce conditions described by eqs.(12,13): $\varepsilon_z\sim$const$<0$ and $\varepsilon_x \sim e^{-Hz}$. Similarities between the layered 3D hyperbolic metamaterials shown in Fig.2(b) and the PMMA-based plasmonic hyperbolic metamaterial (Fig.1b) will allow us to use results of the 3D model described by eqs.(12,13). Let us consider a surface plasmon (SP) wave which propagates over a

flat metal-dielectric interface. If the metal film is thick, the SP wave vector is defined by expression

$$k_p = \frac{\omega}{c}\left(\frac{\varepsilon_d \varepsilon_m}{\varepsilon_d + \varepsilon_m}\right)^{1/2} \qquad (15)$$

where $\varepsilon_m(\omega)$ and $\varepsilon_d(\omega)$ are the frequency-dependent dielectric constants of the metal and dielectric, respectively [15]. Let us introduce an effective 2D dielectric constant $\varepsilon_{2D}$ so that $k_p = \varepsilon_{2D}^{1/2}\omega/c$, and thus

$$\varepsilon_{2D} = \left(\frac{\varepsilon_d \varepsilon_m}{\varepsilon_d + \varepsilon_m}\right) \qquad (16)$$

Now it is easy to see that depending on the frequency, SPs perceive the dielectric material bounding the metal surface in drastically different ways. At low frequencies $\varepsilon_{2D} \approx \varepsilon_d$. Therefore, plasmons perceive a PMMA stripe as dielectric. On the other hand, at high enough frequencies around $\lambda_0 \sim 500$ nm, $\varepsilon_{2D}$ changes sign and becomes negative since $\varepsilon_d(\omega) > -\varepsilon_m(\omega)$. As a result, around $\lambda_0 \sim 500$ nm plasmons perceive PMMA stripes on gold as if they are "metallic layers", while gold/vacuum portions of the interface are perceived as "dielectric layers". Thus, at these frequencies plasmons perceive a PMMA stripe pattern from Fig.1(b) as a layered hyperbolic metamaterial shown in Fig.2(b), and we may use the results of the 3D model of cosmological inflation described by eq.(12,13). Note that rigorous theoretical description of the PMMA-based plasmonic metamaterials developed in ref. [16] produces similar answer. Fabrication of the 2D plasmonic hyperbolic metamaterial requires only very simple and common lithographic techniques. As described in the 3D experimental scenario above, the effective "metallic layer" width $d_1$ of the PMMA stripes was varied, while the width of effective "dielectric layers" (the gold/vacuum portions of the interface) $d_2$ was kept constant. The required concentric structures were defined using a Raith E-line electron beam lithography (EBL) system with ~70 nm spatial resolution. The written structures were subsequently developed using a 3:1 IPA/MIBK solution (Microchem) as developer and imaged using AFM (Fig.2(c)). In order to achieve the required behavior of dielectric permittivity various chirped PMMA ring gratings were created (Fig.2(d,e)). Grating pitch was increased radially with 10 nm increments starting at 300 nm. In the left concentric ring structure in Figs.2(d,e) the ring periodicity is increased according to the (300, 300, 310, 310, 320, 320…) pattern, and in the right structure the pattern is (300, 300, 300, 310, 310, 310…). The fabricated structures were studied using an optical microscope under illumination with P-polarized 488 nm Argon ion laser, as described in [17]. Illumination angle was varied in order to achieve phase-matched excitation of SPs by the concentric ring grating. In Fig.2 the 488 nm illumination was from a 45 degree elevation incident from the right side of the image going across the pattern. The image shown in Fig.2(e) was formed by SP scattering into light and captured by a CCD camera mounted onto the microscope. In agreement with our theoretical design described above, exponentially expanding pattern of ray propagation is indeed clearly visible in both cases. However, the difference between the geometries of left and right patterns in Fig.2e affects the spatial location at which "plasmonic metamaterial space" begins its exponential expansion. This radial location differs by a factor of ~1.5 in agreement with the metamaterial design described above. The logarithmic plot of the "metamaterial space" size $S$ as a function of $r=\tau$ for the left pattern from Fig.2(e), which is shown in Fig.3(b) clearly demonstrates exponential increase of "world line" separation as a function of model time.

11## III. Conclusion.

Thus, a rudimentary metamaterial model of cosmological inflation has been built in our experiments performed with two-dimensional plasmonic hyperbolic metamaterials. These measurements would be much more interesting in the 3D metamaterial model of inflation based on nonlinear layered hyperbolic metamaterial mentioned above, in which inflation will be driven by self-interaction due to nonlinear optical effects.

**Figure Captions**

**Fig. 1**. Experimental demonstration of world lines behavior in an "expanding universe" using a plasmonic hyperbolic metamaterial: (a) Schematic view of world lines behavior near the big bang in the absence of inflation. (b) AFM image of the plasmonic hyperbolic metamaterial based on PMMA stripes on gold. The defect used as a plasmon source is shown by an arrow. (c) Plasmonic rays or "world lines" increase their spatial separation as a function of "timelike" radial coordinate. The point (or moment) $r=\tau=0$ corresponds to a toy "big bang". For the sake of clarity, excitation light scattering by the edges of the PMMA pattern is partially blocked by semi-transparent triangles.

**Fig. 2**. Experimental demonstration of the metamaterial "inflationary universe": (a) Schematic view of world lines behavior near the big bang in the inflationary cosmological theories. Extremely fast exponential expansion of space occurs during a brief period of time immediately after the big bang. (b) Schematic view of the "layered" 3D hyperbolic metamaterial made of subwavelength metal and dielectric layers, which can be used to emulate inflation. (c) Example of an AFM image of the 2D plasmonic metamaterial based on chirped concentric PMMA rings on the gold film surface. (d) Microscope images of two chirped PMMA ring gratings under white light illumination. Grating pitch was increased radially with 10 nm increments starting at 300 nm. In the left concentric ring structure the ring periodicity is increased according to the (300, 300, 310, 310, 320, 320…) pattern, while in the right structure the pattern is (300, 300, 300, 310, 310, 310…). (e) Same metamaterial structures illuminated with 488 nm laser light. Exponentially expanding pattern of ray propagation is clearly visible in both cases.

**Fig. 3**. Magnified view of ray propagation (a), and the logarithmic plot (b) of "metamaterial space" size $S$ as a function of $r=\tau$ clearly demonstrate exponential

increase of "world line" separation as a function of model time in the left image from Fig.2(e).

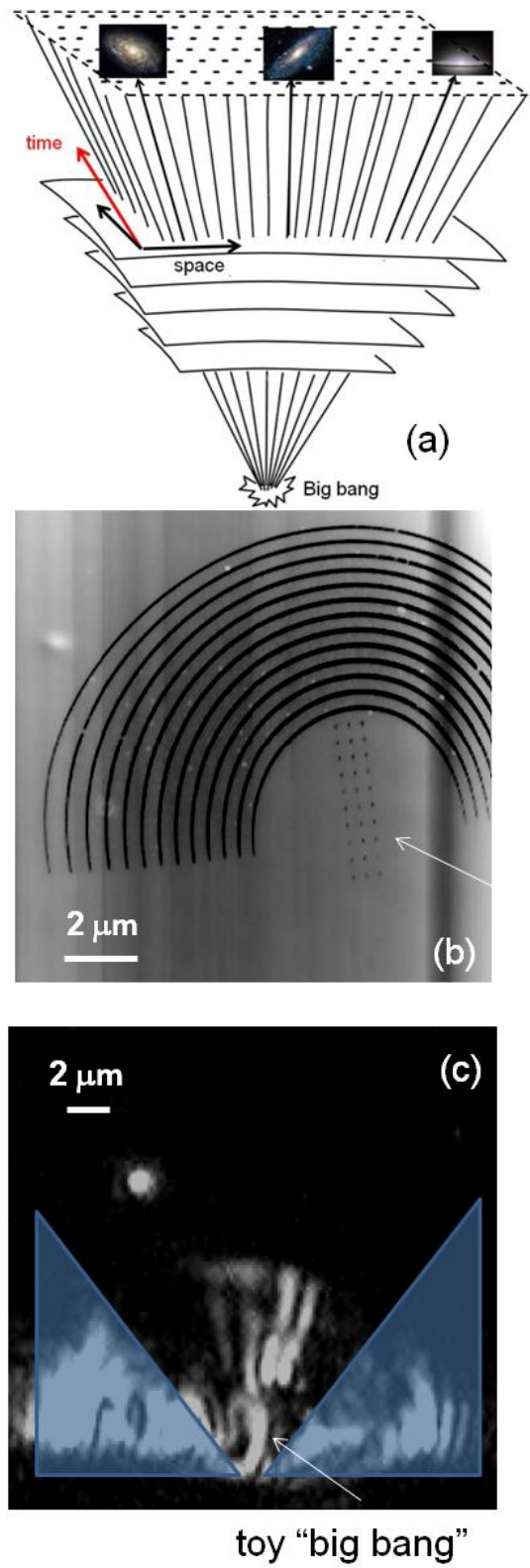

Fig.1

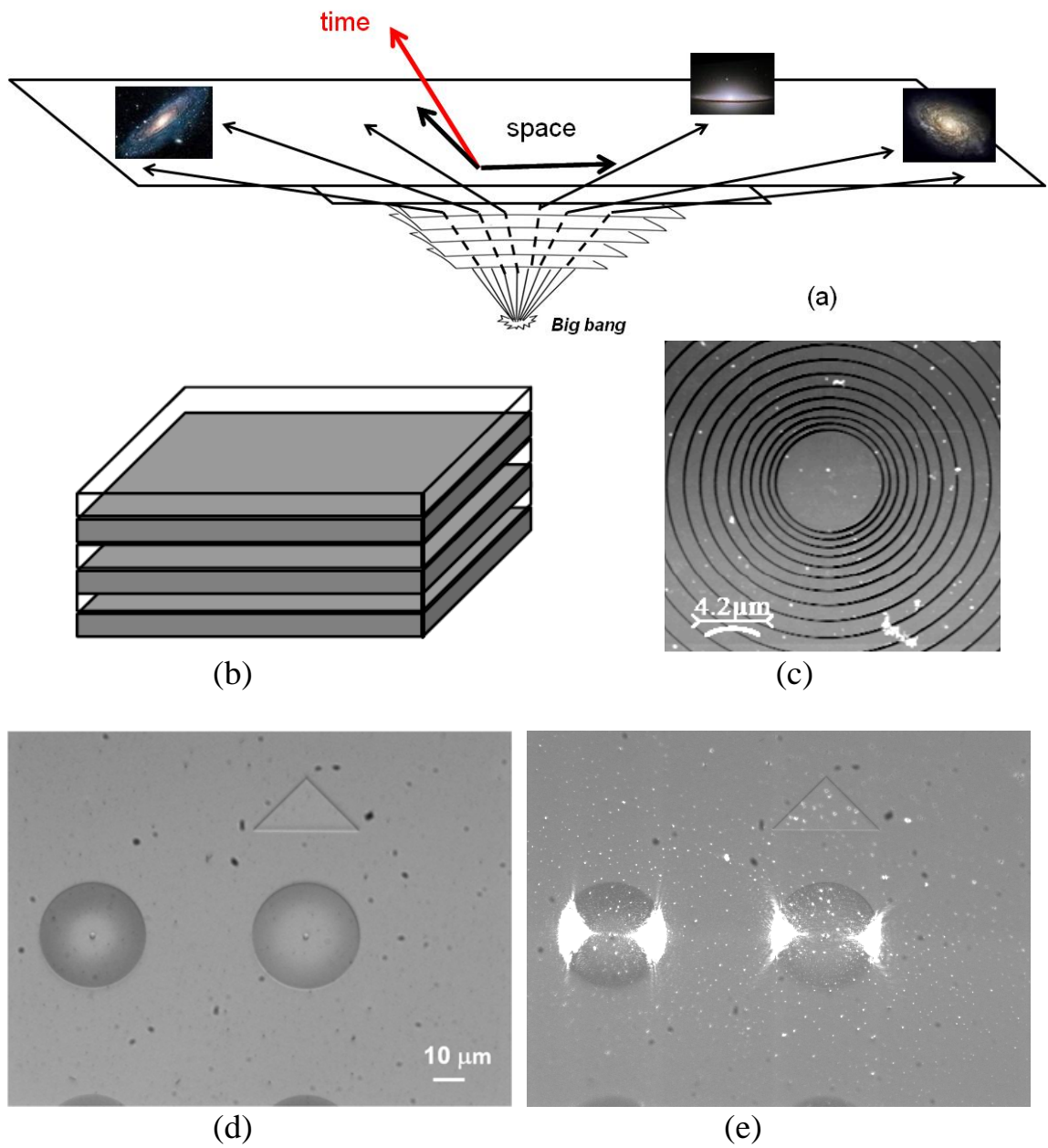

Fig.2

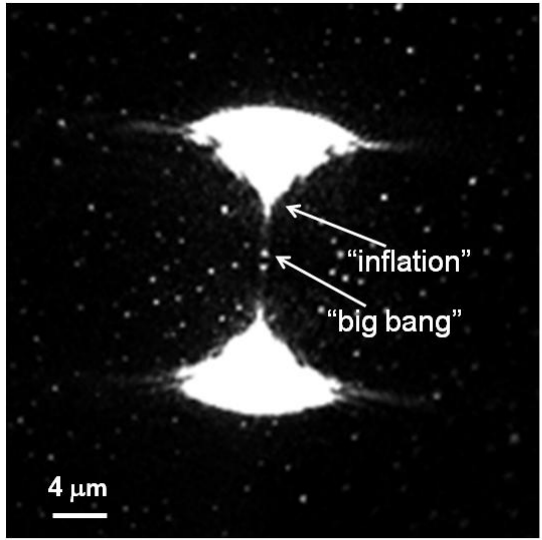
(a)

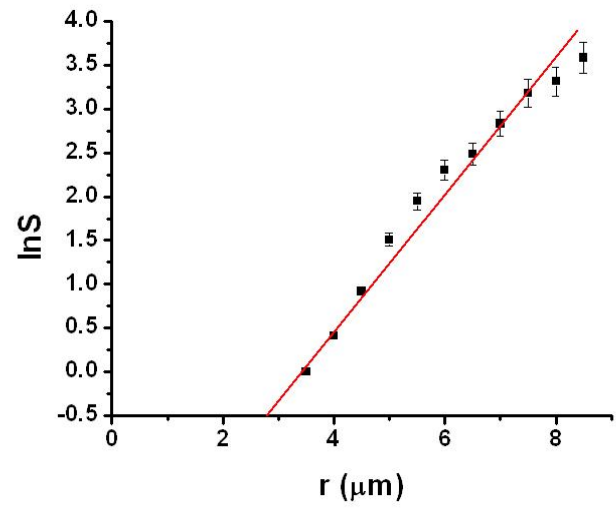
(b)

Fig.3